\documentclass[prd, 11pt,nofootinbib, superscriptaddress, preprintnumbers,floatfix]{revtex4}
\usepackage{amssymb,amsmath}
\usepackage{graphicx}
\usepackage{multirow}
\usepackage{bm}
\usepackage{comment}
\usepackage{color}
\usepackage{epstopdf}

\begin{document}
\newcommand{\ignore}[1]{}
\def\mc#1{{\mathcal #1}}
\newcommand{\be}{\begin{equation}}
\newcommand{\ee}{\end{equation}}
\newcommand{\ba}{\begin{eqnarray}}
\newcommand{\ea}{\end{eqnarray}}
\newcommand{\order}[1]{\mathcal{O}(#1)}
\newcommand{\cir}[1]{\mathring{#1}}

\preprint{JLAB-THY-12-1599}
\title{Interactions of Charmed Mesons with Light Pseudoscalar Mesons from Lattice QCD
and Implications on the Nature of the {\boldmath$D_{s0}^*(2317)$}}

\author{Liuming~Liu}
\email{liuming@maths.tcd.ie}
\affiliation{School of Mathematics, Trinity College Dublin, Ireland}
\affiliation{Helmholz-Institut f\"ur Strahlen- und Kernphysik and Bethe Center for Theoretical
             Physics, Universit\"at Bonn, D-53115 Bonn, Germany}

\author{Kostas~Orginos}
\email{kostas@wm.edu}
\affiliation{Department of Physics, College of William, Williamsburg, VA 23187-8795, USA}
\affiliation{Thomas Jefferson National Accelerator Facility, Newport News, VA 23606, USA}

\author{Feng-Kun~Guo}
\email{fkguo@hiskp.uni-bonn.de}
\affiliation{Helmholz-Institut f\"ur Strahlen- und Kernphysik and Bethe Center for Theoretical
             Physics, Universit\"at Bonn, D-53115 Bonn, Germany}

\author{Christoph~Hanhart}
\email{c.hanhart@fz-juelich.de}
\affiliation{Institute for Advanced Simulation, Institut f\"{u}r Kernphysik
             and J\"ulich Center for Hadron Physics, 
             Forschungszentrum J\"ulich, D-52425 J\"{u}lich, Germany}

\author{Ulf-G.~Mei{\ss}ner}
\email{meissner@hiskp.uni-bonn.de}
\affiliation{Helmholz-Institut f\"ur Strahlen- und Kernphysik and Bethe Center for Theoretical
             Physics, Universit\"at Bonn, D-53115 Bonn, Germany}
\affiliation{Institute for Advanced Simulation, Institut f\"{u}r Kernphysik, 
             J\"ulich Center for Hadron Physics and JARA -- High Performance
             Computing, Forschungszentrum J\"ulich, D-52425 J\"{u}lich, Germany}


\begin{abstract}
We study the scattering of light pseudoscalar mesons ($\pi$, $K$) off charmed mesons 
($D$, $D_s$) in full lattice QCD. The $S$-wave scattering  lengths are
calculated using L\"uscher's finite volume technique. We use a
relativistic formulation for the charm quark. For the light quark,  we
use domain-wall fermions in the valence sector and improved Kogut-Susskind sea 
quarks. We calculate the scattering lengths of isospin-3/2 $D\pi$, $D_s\pi$,
$D_sK$,  isospin-0 $D\bar{K}$ and isospin-1 $D\bar{K}$ channels on the lattice.  
For the chiral extrapolation, we use a chiral unitary approach to next-to-leading 
order, which at the same time allows us to give predictions for other channels. 
It turns out that our results support the interpretation of the $D_{s0}^*(2317)$ as 
a $DK$ molecule. At the same time, we also update a prediction for the isospin breaking 
hadronic decay width $\Gamma(D_{s0}^*(2317)\to D_s\pi)$ to $(133\pm19)$~keV. 
\end{abstract}

\maketitle

\newpage

\section{Introduction}

In 2003, the BaBar Collaboration discovered a positive-parity scalar charm-strange 
meson $D_{s0}^*(2317)$ with a very narrow width~\cite{Aubert:2003fg}. The state was 
confirmed later by the CLEO
Collaboration~\cite{Besson:2003cp}. The discovery of
this state has inspired heated discussions in the past decade. The key
point is to understand the low mass of this state, which is more than 100~MeV lower
than the prediction for the lowest scalar $c\bar s$ meson in, for instance, the
Godfrey-Isgur quark model~\cite{Godfrey:1985xj}.
There are several interpretations of its structure, such as being a $DK$ molecule, the chiral 
partner of the pseudoscalar $D_s$, a conventional $c\bar{s}$ state, coupled-channel
effects between the $c\bar{s}$ state and $DK$ continuum etc. For a detailed review of the
properties and the phenomenology of these states see Ref.~\cite{Zhu:2007wz}. 
In order to distinguish them, one has to explore the consequences of 
each interpretation, and identify quantities which have different values in different 
interpretations. Arguably the
most promising quantity is the isospin breaking width $\Gamma(D_{s0}^*(2317)\to
D_s \pi)$. It is of order 10~keV if the $D_{s0}^*(2317)$ is a $c\bar s$
meson~\cite{Godfrey:2003kg,Colangelo:2003vg}, while it is of order
100~keV~\cite{Faessler:2007gv,Lutz:2007sk,Guo:2008gp} in the $DK$ molecular picture
due to its large coupling to $DK$ and the proximity of the $DK$ threshold. Thus, the
study of $DK$ interaction is very important in order to understand the structure of
$D_{s0}^*(2317)$ 
(a suggestion of studying this state in a finite volume was made in Ref.~\cite{MartinezTorres:2011pr}).
Although a direct simulation of the $DK(I=0)$ channel suffers from disconnected
diagrams, one may obtain useful information on the $DK$ interaction by calculating
the scattering lengths of the disconnected-diagram-free channels which can be related
to $DK(I=0)$ through SU(3) flavor symmetry. This is the strategy we will follow in
this paper.

Lattice QCD calculations of  the properties of hadronic interactions such as
elastic  scattering phases shifts and scattering lengths have recently started
to develop. Precision results have been obtained in the  light meson sector for
certain processes such as pion-pion, kaon-kaon and pion-kaon scattering and 
preliminary  results for baryon-baryon scattering lengths have been presented.   
A review of these calculations can be found in Ref.~\cite{Beane:2008dv}. In the heavy
meson sector, only a few calculations have been 
done, including quenched calculations in Refs.~\cite{Yokokawa:2006td,Meng:2009qt}
and calculations in full QCD in Refs.~\cite{Liu:2008rza,Mohler:2012}.
In this work, we study scattering processes  where one of the hadrons contains a charm quark in full lattice QCD.

Extracting hadronic interactions from Lattice QCD calculations is not straightforward
due to the Maiani-Testa theorem~\cite{Maiani:1990ca}, which states that the
$S$-matrix can not be extracted from infinite-volume Euclidean-space Green functions except at 
kinematic thresholds. However, this problem can be evaded by computing the
correlation functions in a finite volume.  L\"uscher has shown that one can 
obtain the scattering amplitude from the energy of two particles in a finite 
volume~\cite{Luscher:1986pf, Luscher:1990ux}. We use L\"uscher's finite volume
technique to calculate the scattering lengths. We then use unitarized 
chiral perturbation theory to extrapolate our results to the physical pion
mass. Having fitted the appearing low-energy constants (LECs) to the lattice data,
we are also able to make predictions for other channels, in particular for
the isospin zero, strangeness one channel in which the $D_{s0}^*(2317)$ resides. 

The paper is organized as follows. The lattice formulation of the light and heavy
quark actions will be discussed in Section~\ref{sec:LatticeAction}. L\"uscher's
formula will be briefly introduced in Section~\ref{sec:luescher}. The numerical
results for the scattering lengths of five channels $D\bar K(I=0)$, $D\bar K(I=1)$,
$D_s K$, $D\pi(I=3/2)$ and $D_s\pi$ which are free of disconnected diagrams will be
given in Section~\ref{sec:numerical}. Chiral extrapolations will be performed using
unitarized chiral perturbation theory in Section~\ref{sec:chiralex}, and values of
the LECs in the chiral Lagrangian will be determined.
Predictions for other channels using the LECs are given Section~\ref{sec:ds2317}, and
in particular, implications on the $D_{s0}^*(2317)$ will be discussed. The last
section is devoted to a brief summary.

\section{Lattice Formulation }
\label{sec:LatticeAction}

\subsection{Light-Quark Action}
\label{sec:lQaction}

In this work we employ the ``coarse'' ($a\simeq 0.125$~fm) gauge configurations 
generated by the MILC Collaboration~\cite{Bernard:2001av} using the one-loop 
tadpole-improved gauge action~\cite{Alford:1995hw}, where both $\mc{O}(a^2)$ and 
$\mc{O}(g^2a^2)$ errors are removed. For the fermions in the vacuum, the 
asqtad-improved Kogut-Susskind (staggered)
action~\cite{Orginos:1999cr,Orginos:1998ue,Toussaint:1998sa,Lagae:1998pe,Lepage:1998vj,Orginos:1999kg}
is used. This is the so-called Naik action~\cite{Naik:1986bn} ($\mc{O}(a^2)$ improved 
Kogut-Susskind action) with smeared links for the one-link terms so that  couplings
to gluons with any of their momentum components equal to $\pi/a$ are set to zero,
resulting in a reduction of the flavor symmetry violations present in the
Kogut-Susskind action.

For the valence light quarks (up, down and strange) we use the five-dimensional 
Shamir~\cite{Shamir:1993zy,Furman:1994ky} domain-wall fermion 
action~\cite{Kaplan:1992bt}.  The domain-wall fermion action introduces a fifth
dimension of extent $L_5$ and a mass parameter $M_5$; in our case, the values
$L_5=16$ and $M_5=1.7$, both in lattice units, were chosen. The physical
quark fields, $q(\vec x, t)$, reside on the 4-dimensional boundaries 
of the fifth coordinate. The left and right chiral
components  are separated on the corresponding boundaries, resulting in an action 
with chiral symmetry at finite lattice spacing as $L_5 \rightarrow \infty$. We use 
hypercubic-smeared gauge
links~\cite{Hasenfratz:2001hp,DeGrand:2002vu,DeGrand:2003in,Durr:2004as}  to minimize
the residual chiral symmetry breaking, and the bare quark-mass parameter $(a m)^{\rm
dwf}_q$  is introduced as a direct coupling of the boundary chiral components. The
light quark propagators were provided to us by the NPLQCD~\cite{Beane:2008dv} and
LHP~\cite{Renner:2004ck,Edwards:2005kw,WalkerLoud:2008bp} Collaborations.

The calculation we have performed, because the valence and sea quark actions are
different,  is inherently partially quenched and therefore violates unitarity. 
Unlike  conventional partially quenched calculations, to restore unitarity, one must 
take the continuum limit in addition to tuning the valence and sea quark masses to be
degenerate.  This process is aided by the use of mixed-action chiral perturbation 
theory~\cite{Bar:2005tu,Tiburzi:2005is,Chen:2006wf,Orginos:2007tw,Chen:2007ug,Chen:2009su}.
Given the situation, there is an ambiguity in the choice of the valence light-quark
masses. One appealing choice is to tune the valence light quark masses such that the
valence pion mass is degenerate with the Goldstone staggered pion mass.  
In the continuum limit, the 
$N_f=2$ staggered action has an $SU(8)_L\otimes SU(8)_R\otimes U(1)_V$ chiral
symmetry  due to the four-fold taste degeneracy of each flavor, and each pion has  15
degenerate partners. At finite lattice spacing this symmetry is broken and the taste 
multiplets are no longer degenerate, but have splittings that are 
$\mc{O}(\alpha_s^2 a^2)$~\cite{Orginos:1999cr,Orginos:1998ue,Toussaint:1998sa,Orginos:1999kg,Lee:1999zxa}.
The propagators used in this work were tuned to give valence pions that match the 
Goldstone Kogut-Susskind pion. This is the only pion that becomes massless in the 
chiral limit at finite lattice spacing. As a result of this choice, the valence pions
are as light as possible, while being tuned to one of the staggered pion masses, 
providing better convergence in the chiral perturbation theory needed to extrapolate
the lattice results to the physical quark-mass point.  This set of parameters, listed in 
Table~\ref{table:configuration}, was first used by
LHPC~\cite{Renner:2004ck,Edwards:2005kw}  and utilized to compute the
spectroscopy of hadrons  composed of up, down and strange quarks~\cite{WalkerLoud:2008bp}. A two-flavor chiral perturbation theory analysis on this action was recently performed for the pion mass and pion decay constant~\cite{Beane:2011zm}, finding good agreement with the lattice average of these quantities and their LEC's.

\begin{table}
\begin{ruledtabular}
\begin{tabular}{|l|ccccccc|}
Ensemble  &$\beta$ &$am_l$  &$am_s$ & $am_l^{\rm dwf}$ & $am_s^{\rm dwf}$ & $N_{\rm cfgs}$ &$N_{\rm props}$ \\
\hline
\texttt{m007}& 6.76   & 0.007    & 0.050   & 0.0081   & 0.081    & 461   & 2766 \\
\texttt{m010}& 6.76   & 0.010    & 0.050   & 0.0138   & 0.081    & 636   & 3816 \\
\texttt{m020}& 6.79   & 0.020    & 0.050   & 0.0313   & 0.081    & 480   & 1920 \\
\texttt{m030}& 6.81   & 0.030    & 0.050   & 0.0478   & 0.081    & 563   & 1689 \\
\end{tabular}
\end{ruledtabular}
\caption{\label{table:configuration}The parameters of the configurations and
domain-wall  propagators used in this work. The subscript $l$ denotes light quark,
and  $s$ denotes the strange quark. The superscript ``dwf'' denotes domain-wall fermion.}
\end{table}

%
%
\subsection{Heavy-Quark Action}
\label{sec:HQaction}

For the charm quark we use a relativistic heavy quark action motivated by the
Fermilab approach~\cite{ElKhadra:1996mp}. This action controls discretization
errors of $\mc{O}((a m_Q)^n)$.
Following the Symanzik improvement~\cite{Symanzik:1983dc}, an effective continuum
action  is constructed using operators that are invariant under discrete rotations, 
parity-reversal and charge-conjugation transformations, representing the
long-distance  limit of our lattice theory, including leading finite-$a$ errors.
Using only the Dirac operator and the gluon field tensor (and distinguishing between 
the time and space components of each), we enumerate seven operators with dimension up to five.
By applying the isospectral transformations~\cite{Chen:2000ej}, the redundant
operators  are identified and their coefficients are set to appropriate  convenient values.
The lattice action then takes the form
\begin{equation}\label{eq:SFermi}
	S = S_0+S_B+S_E\, ,
\end{equation}
with
\begin{align}
\label{eq:action_S_0}
&S_0 = \sum_x \bar{Q}(x)\left[m_0+\left(\gamma_0 \nabla_0-\frac{a}{2}\triangle_0\right)
	+ \nu \sum_i\left(\gamma_i\nabla_i-\frac{a}{2}\triangle_i\right)
	\right]Q(x)\, ,& \\
\label{eq:action_S_B}
&S_B = -\frac{a}{2}c_B\sum_x \bar{Q}(x)\left(\sum_{i<j}\sigma_{ij}F_{ij} \right) Q(x)\, ,&\\
\label{eq:action_S_E}
&S_E = -\frac{a}{2}c_E\sum_x \bar{Q}(x) \left(\sum_i\sigma_{0i}F_{0i} \right) Q(x)\, ,&
\end{align}
where the operator $Q(x)$ annihilates a heavy quark field, $a$ is the lattice
spacing, $\nabla_0$ and $\nabla_i$ are first-order lattice derivatives in the time and space directions, $\triangle_0$ and $\triangle_i$ are
second-order  lattice derivatives, and $F_{\mu\nu}$ is the gauge field strength tensor.
The spectrum of heavy-quark bound states can be determined accurately through
$|\vec{p}|a$  and $(am_Q)^n$ for arbitrary exponent $n$ by using a lattice action 
containing $m_0$, $\nu$, $c_B$ and $c_E$, which are functions of $am_Q$.

The coefficients $c_B$ and $c_E$ are different due to the broken space-time
interchange  symmetry, which can be computed in perturbation theory by requiring 
elimination of the heavy-quark discretization errors at a given order in the strong 
coupling constant $\alpha_s$. We use the tree-level tadpole-improved results obtained
by using field transformation (as in Ref.~\cite{Chen:2000ej}):
\begin{equation}\label{eq:cBcE}
	c_B=\frac{\nu}{u_0^3}, \quad \quad
	c_E=\frac{1}{2}(1+\nu)\frac{1}{u_0^3} ,
\end{equation}
where $u_0$ is the tadpole factor
\begin{equation}
	u_0 = \left \langle \frac{1}{3} \sum_p {\rm Tr} (U_p )\right\rangle^{1/4}\;,
\end{equation}
and $U_p$ is the product of gauge links around the fundamental lattice plaquette $p$.
The remaining two parameters $m_0$ and $\nu$ are determined nonperturbatively. The
bare  charm-quark mass $m_0$ is tuned so that the experimentally observed spin
average  of the $J/\psi$ and $\eta_c$ masses
\begin{equation}
	M_{\rm avg} = \frac{1}{4} M_{\eta_c} + \frac{3}{4} M_{J/\psi}
\label{eq:spin-av}
\end{equation}
is reproduced. For each ensemble, we calculate $M_{\rm avg}$ at two charm-quark masses (denoted $m_1 = 0.2034$ and $m_2 = 0.2100$) and linearly extrapolate it to the experimental value to determine the parameter $m_0 = m_c^{\rm phys}$. The value of 
$\nu$ must be tuned to restore the dispersion relation $E_h^2=m_h^2+c^2p^2$ such 
that $c^2 = 1$. To do this, we calculate the single-particle energy of $\eta_c$,
$J/\psi$, $D_s$ and $D$ at the six lowest momenta (with unit of $a^{-1}$): 
$(2\pi/L)(0,0,0), (2\pi/L)(1,0,0),$ 
$(2\pi/L)(1,1,0),(2\pi/L)(1,1,1), (2\pi/L)(2,0,0), (2\pi/L)(2,1,0)$. 
For each ensemble, the energy levels are calculated at the two charm-quark masses 
($m_1$ and $m_2$) and extrapolated to the physical charm-quark mass $m_c^{\rm phys}$. 
The values of $c^2$ are obtained by fitting the extrapolated energy levels to the 
dispersion relation. We tune $\nu$ using the dispersion relation of $\eta_c$. The
dispersion relations for either the charmonium $J/\psi$ or the charm-light mesons 
($D$ and $D_s$) are generally consistent with $c^2 = 1$ within 1-2\%.
 Since the values of $\nu$ and $m_0$ are coupled, one needs to
iterate  the tuning process in order to achieve a consistent pair of values.  For the
details of tuning the bare charm-quark mass $m_0$ and the value of $\nu$, see 
reference~\cite{Liu:2009jc}.

\section{ L\"{u}scher's formula}
\label{sec:luescher}

L\"{u}scher has shown
that the scattering phase shift is related to the energy shift
($\Delta E$) in the total energy of two interacting hadrons in a finite 
box~\cite{Luscher:1986pf, Luscher:1990ux}.

The center-of-mass momentum $p$ can be obtained by the relation
\begin{equation}
\label{Eq:DeltaE}
\Delta E=\sqrt{m_1^2+p^2}+\sqrt{m_2^2+p^2}-m_1-m_2,
\end{equation}
where $m_1$ and $m_2$ are the rest masses of the two hadrons.

To obtain $p\cot \delta(p)$, where $\delta(p)$ is the phase shift, we use the 
formula~\cite{Beane:2003da}
\begin{equation}
 \label{Eq:pcotdelta1}
p\cot \delta(p) = \frac{1}{\pi
L}\textbf{S}\Big(\Big(\frac{pL}{2\pi}\Big)^2\Big) ,
\end{equation}
  where the $\textbf{S}$ function is defined as
  \begin{equation}
  \label{eq:Sfunc}
  \textbf{S}(x)=\sum_{\textbf{j}}^{|\textbf{j}| < \Lambda}
  \frac{1}{|\textbf{j}|^2-x}-4\pi\Lambda .
  \end{equation}
  The sum is over all three-vectors of integers $\textbf{j}$ such
  that $|\textbf{j}|< \Lambda$, and the limit $\Lambda \rightarrow \infty$ is implicit.
  
   If the interaction range
  is much smaller than the lattice size, $p\cot \delta(p)$ 
  is given by
  \begin{equation}
  \label{Eq:pcotdelta2}
  p\cot \delta(p) = \frac{1}{a} + \mathcal {O}
  (p^2) ,
  \end{equation}
  where $a$ is the scattering length (not to be confused with the lattice spacing
  which has the same notation and dimension). Note that we take the sign convention
  that a repulsive interaction has a negative scattering length. The higher order
  terms in Eq.~(\ref{Eq:pcotdelta2}) can be ignored if the effective range of the
  interaction is much smaller than the length scale associated to the center-of-mass 
  momentum $p$. If we ignore the higher order terms, the scattering length can be calculated by
  \begin{equation}\label{Eq:ScattLen}
  a=  \pi L \textbf{S}^{-1} \Big(\Big(\frac{pL}{2\pi}\Big)^2\Big).
\end{equation}

\section{Numerical Results}
\label{sec:numerical}

In the following, we list all the channels we study. The interpolating operators for
these  two particle states are
\begin{eqnarray*}
&&\mathcal{O}_{D_s\pi}   = D_s^- \pi^+ , \quad
\mathcal{O}_{D\pi}^{I=3/2} = D^+\pi^+ , \quad \mathcal{O}_{D_sK} = D_s^+K^+ ,\\
&&\mathcal{O}_{D\bar{K}}^{I=1} =D^+ \bar{K}^0 , \quad \mathcal{O}_{D\bar{K}}^{I=0} 
=D^+K^- - D^0 \bar{K^0},
\end{eqnarray*}
where $D_s^-$, $D_s^+$, $D^+$, $K^0$, $K^-$, $K^+$ and $\pi^+$ are the operators for 
one particle states, the subscripts $\pi$, $D$, $K$ and $\bar{K}$ represent the
isospin  triplet $(\pi^+$, $\pi^0$, $\pi^-$) and doublets ($D^+$, $D^0$), ($K^+$,
$K^0$)  and ($\bar{K}^0$, $K^-$), respectively.

 The total
energy of two interacting hadrons ($h_1$ and $h_2$) is obtained from
the four-point correlation function:
\begin{equation}
G^{h_1h_2}(t)=\langle \mathcal{O}_{h_1h_2}(t)^\dag
\mathcal{O}_{h_1h_2}(0)\rangle .
\end{equation}
To extract the energy shift $\triangle E$, we define a ratio
$R^{h_1h_2}(t)$:
\begin{equation}
R^{h_1h_2}(t)=\frac{G^{h_1h_2}(t)}{G^{h_1}(t)G^{h_2}(t)}\longrightarrow
\exp(-\triangle E \cdot t) ,
\end{equation}
where $G^{h_1}(t,0)$ and $G^{h_2}(t,0)$ are two-point functions.
$\triangle E$ is obtained by fitting $R^{h_1h_2}(t)$ to a
single exponential in the region where the effective mass exhibits a plateau.

\begin{figure}[tbh]
\begin{center}
\includegraphics[width = 0.495\textwidth]{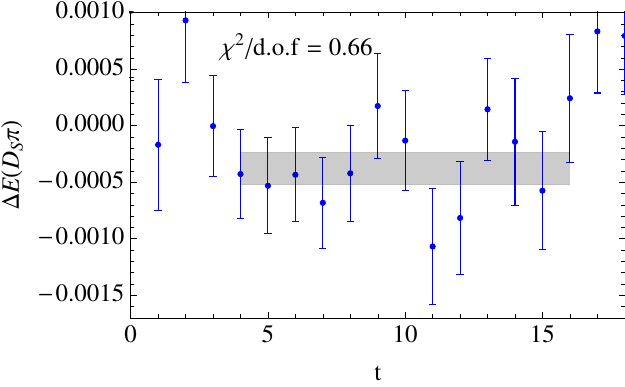}\hfill
\includegraphics[width = 0.495\textwidth]{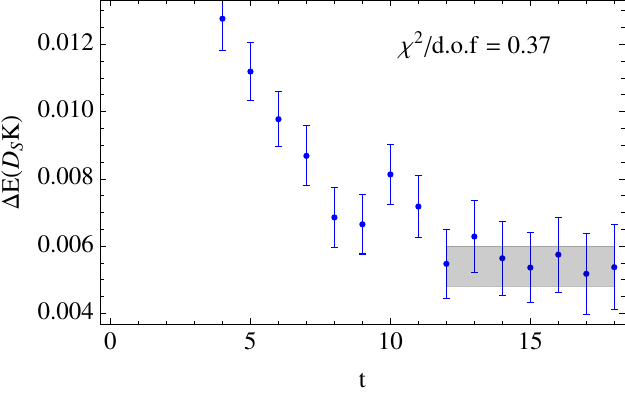}\\
\includegraphics[width = 0.495\textwidth]{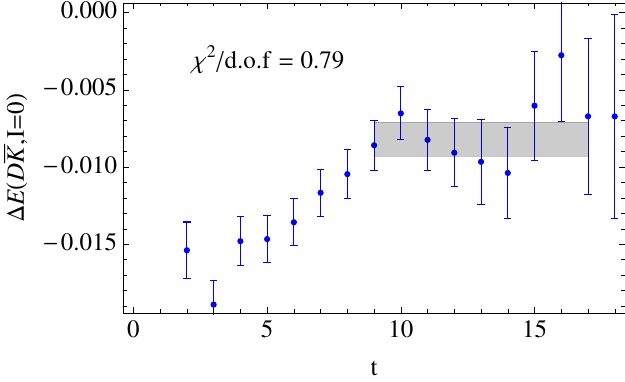}\hfill
\includegraphics[width = 0.495\textwidth]{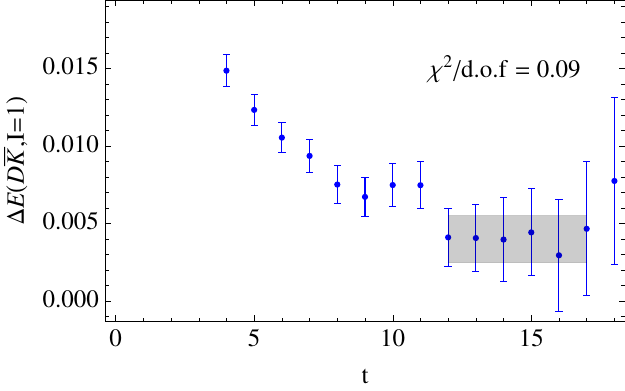}\\
\includegraphics[width = 0.495\textwidth]{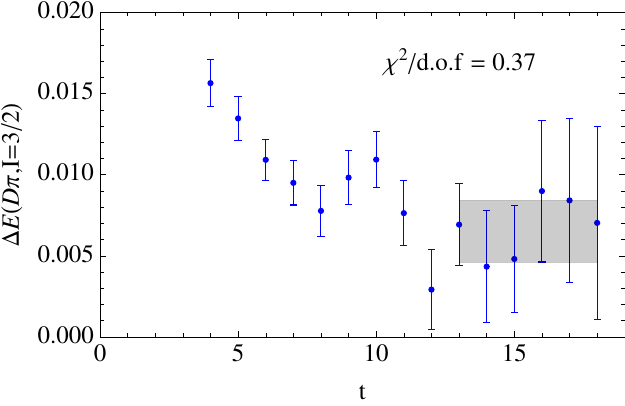}
\end{center}
\caption{\label{Fig:EffectiveEnergyShift} Effective energy shifts plots of the
scattering  channels $D_s\pi$, $D_sK$, $D\bar{K}(I=0)$, $D\bar{K}(I=1)$,
$D\pi(I=3/2)$.  All plots are for ensemble \texttt{m007}. The grey bars show the
fitted  energy shifts and the fitting ranges. The height of the grey bars show the 
statistical errors.}
\end{figure}

For each channel, we calculate the ratio $R^{h_1h_2}$ at two different charm quark
masses  and four different light valence quark masses.
Figure~\ref{Fig:EffectiveEnergyShift}  shows the effective energy shifts of each
channel  calculated from ensemble $\texttt{m007}$ at the bare charm-quark mass 
$m_2 = 0.2100$. The fitted energy shifts and the
fitting  ranges are indicated by the grey bars in these plots. The heights of the grey
bars  show the statistical errors. The $\chi^2$ per degree of freedom for these fits are presented in the plots. The fits of the energy shifts for other
ensembles  are similar.

The energy shifts are linearly extrapolated to the physical charm-quark mass on each ensemble. 

\section{Chiral extrapolations of the scattering lengths}
\label{sec:chiralex}

Because the simulations are performed at unphysical quark masses, chiral
extrapolation is necessary in order to obtain the values of scattering lengths
at the physical quark masses. There have been calculations based on chiral
Lagrangians for these scattering
lengths~\cite{Guo:2009ct,Liu:2009uz,Geng:2010vw,Wang:2012bu}. They were first calculated in
Ref.~\cite{Guo:2009ct} using a unitarized chiral approach.
The basic observation is because of the coupled-channel effect and the large kaon
mass, the interaction of some of the channels is so strong that a
nonperturbative treatment is necessary, and in one channel even a bound state is produced. 
The method was followed up recently in Ref.~\cite{Wang:2012bu}. Some other authors 
treated the interaction perturbatively, and calculated the scattering lengths up to 
leading one-loop order in chiral perturbation theory with~\cite{Liu:2009uz} and 
without~\cite{Geng:2010vw} a heavy quark expansion. Here we take the same
route as Ref.~\cite{Guo:2009ct}, and resum the chiral amplitude up to next-to-leading order, which is
$\order{p^2}$. The resummed amplitude in the on-shell approximation
reads~\cite{Oller:1997ti,Oller:1998zr,Oller:2000fj}
\be
\label{eq:t}
T(s) = V(s) [1-G(s) V(s)]^{-1},
\ee
where $V(s)$ is the $S$-wave projection of the $\order{p^2}$ scattering amplitude,
and $G(s)$ is the scalar loop function regularized by a subtraction constant
$\tilde{a}(\lambda)$
\ba
G(s) &\!=&\! \frac{1}{16\pi^2}\bigg\{\tilde{a}(\lambda)+\ln{\frac{m_2^2}{\lambda^2}}
+
\frac{m_1^2-m_2^2+s}{2s}\ln{\frac{m_1^2}{m_2^2}}
+\frac{\sigma}{2s}\left[\ln({s-m_1^2+m_2^2+\sigma}) \right. \nonumber\\
& & \left. -\ln({-s+m_1^2-m_2^2+\sigma}) +
\ln({s+m_1^2-m_2^2+\sigma})-\ln({-s-m_1^2+m_2^2+\sigma}) \right]
\bigg\},
\ea
with $\sigma=\left\{[s-(m_1+m_2)^2][s-(m_1-m_2)^2]\right\}^{1/2}$. $\lambda$ is
the scale of dimensional regularization, and a change of $\lambda$ can be absorbed by
a corresponding change of $\tilde{a}(\lambda)$. The value $\lambda=1$~GeV will be
taken in the following.
Promoting $T(s)$, $V(s)$ and $G(s)$ to be matrix-valued quantities, it is easy to 
generalize Eq.~(\ref{eq:t}) to coupled channels.

\begin{table}
\begin{ruledtabular}
\begin{tabular}{|l c | c c c c c |}
 $(S,I)$ & Channels & $C_{\rm LO}$ & $C_0$ & $C_1$ & $C_{24}$ & $C_{35}$
 \\
\hline
$(-1,0)$      & $D\bar{K}\to D\bar K$   & $-1$  & $M_K^2$ & $M_K^2$  & 1 & $-1$
\\
$(-1,1)$      & $D\bar{K}\to D\bar K$   & 1     & $M_K^2$ & $-M_K^2$ & 1 & 1 
\\
$(2,\frac12)$ & $D_sK\to D_sK$          & 1     & $M_K^2$ & $-M_K^2$ & 1 & 1 
\\
$(0,\frac32)$ & $D\pi\to D\pi$          & 1 & $M_\pi^2$ & $-M_\pi^2$ & 1 & 1
\\
$(1,1)$       & $D_s\pi\to D_s\pi$      & 0   & $M_\pi^2$ & 0        & 1 & 0 
\\
              & $D K\to D K$            & 0     & $M_K^2$ & 0        & 1 & 0 
\\
              & $D K\to D_s\pi$   & 1 & 0 & $-(M_K^2+M_\pi^2)/2$ & 0    & 1
\\
$(1,0)$       & $D K\to D K$            & $-2$ & $M_K^2$ & $-2M_K^2$ & 1 & 2
\\
              & $D_s\eta\to D_s\eta$ & 0 & $M_\eta^2$ & $-2M_\eta^2+2M_\pi^2/3$
             & 1 & $\frac43$
\\
             & $D K\to D_s\eta$ & $-\sqrt{3}$ & 0 &
        $-\sqrt{3}(5M_K^2-3M_\pi^2)/6$ & 0 & $\frac1{\sqrt{3}}$ 
\\  
$(0,\frac12)$ & $D\pi\to D\pi$       & $-2$    & $M_\pi^2$ & $-M_\pi^2$ & 1 & 1
\\
             & $D\eta\to D\eta$     & 0 & $M_\eta^2$& $-M_\pi^2/3$& 1
             & $\frac13$ 
\\
             & $D_s\bar K\to D_s\bar K$& $-1$& $M_K^2$& $-M_K^2$& 1 & 1 
\\ 
             & $D\eta\to D\pi$     & 0 & 0 & $-M_\pi^2$ & 0 & 1 
\\
             & $D_s\bar K\to D\pi$ & $-\frac{\sqrt{6}}{2}$ & 0 &
             $-{\sqrt{6}}(M_K^2+M_\pi^2)/4$ & 0 & $\frac{\sqrt{6}}{2}$
\\
             & $D_s\bar K\to D\eta$& $-\frac{\sqrt{6}}{2}$ & 0 &
             ${\sqrt{6}}(5M_K^2-3M_\pi^2)/12$ & 0 & $-\frac1{\sqrt{6}}$
\\
\end{tabular}
\end{ruledtabular}
\caption{\label{tab:ci}The coefficients in the scattering amplitudes $V(s,t,u)$. The
channels are labelled by strangeness ($S$) and isospin ($I$). }
\end{table}
Using the $\order{p^2}$ chiral Lagrangian constructed in
Ref.~\cite{Guo:2008gp}, the scattering amplitudes are given by
\be
\label{eq:v}
V(s,t,u) = \frac1{F^2} \bigg[\frac{C_{\rm LO}}{4}(s-u) - 4 C_0 h_0 +
2 C_1 h_1 - 2C_{24} H_{24}(s,t,u) + 2C_{35} H_{35}(s,t,u) \bigg],
\ee
where $F$ is the pion decay constant in the chiral limit, and the coefficients $C_i$
can be found in Table~\ref{tab:ci}. Further,
$$
H_{24}(s,t,u) = 2 h_2 p_2\cdot p_4 + h_4 (p_1\cdot p_2 p_3\cdot p_4 +
p_1\cdot p_4 p_2\cdot p_3), 
$$
and 
$$
H_{35}(s,t,u) = h_3 p_2\cdot p_4 + h_5
(p_1\cdot p_2 p_3\cdot p_4 + p_1\cdot p_4 p_2\cdot p_3). 
$$
Note that the term $h_1\tilde\chi_+=h_1(\chi_+-\langle\chi_+\rangle/3)$ in the
Lagrangian in Refs.~\cite{Guo:2008gp,Guo:2009ct} has been replaced by $h_1\chi_+$,
which amounts to a redefinition of $h_0$ (for the details of the Lagrangian and the
definition of $\chi_+$, we refer to Refs.~\cite{Guo:2008gp,Guo:2009ct}). In this way, 
the $h_1$ term does not contain the $1/N_c$, with $N_c$ being the number of colors,
suppressed part $\langle\chi_+\rangle$ any more. This was also done in
Ref.~\cite{Wang:2012bu}.

In previous works~\cite{Guo:2008gp,Guo:2009ct}, the large-$N_c$
suppressed low-energy constants (LECs) $h_{0,2,4}$ were dropped to reduce the number of parameters. However, 
when fitting to the lattice data at several unphysical quark masses, this is no longer
necessary. In this work, we will keep all of the LECs 
at this order, and fit them to the lattice data. This strategy were also taken in
Refs.~\cite{Liu:2009uz,Geng:2010vw,Wang:2012bu}, where the preliminary
lattice results~\cite{Liu:2008rza} were used.
By definition, the LECs are independent of the pion mass. We further need to assume 
that the subtraction constant
is the same for various channels, and neglect its pion mass dependence. In principle,
this assumption is not necessary for a unitarization procedure matched to the full
one-loop level of the perturbative calculation~\cite{Oller:2000fj,Hanhart:2008mx},
which will be left for the future.

From the SU(3) mass splitting of the charmed mesons, the value of $h_1$ is fixed to
be $h_1=0.42$. We still have six parameters, which are $\tilde{a}, h_3, h_5, h_0,
h_2$ and $h_4$. They are to be fitted to the lattice data. However, there is a high
correlation between $h_3$ and $h_5$, as well as a similar correlation between $h_2$ and $h_4$. In
the heavy quark limit, the $S$-wave projected amplitudes cannot distinguish the
$h_{4(5)}$ terms from the $h_{2(3)}$ ones~\cite{Cleven:2010aw}. Hence, we may
reduce the correlations largely by rewriting $H_{24}(s,t,u)$ and $H_{35}(s,t,u)$
as $$
H_{24}(s,t,u) = 2 h_{24} p_2\cdot p_4 + h_4 \left(p_1\cdot p_2 p_3\cdot p_4 +
p_1\cdot p_4 p_2\cdot p_3 - 2 \bar M_D^2 p_2\cdot p_4 \right), 
$$
and
$$
H_{35}(s,t,u) = h_{35} p_2\cdot p_4 + h_5 \left(p_1\cdot p_2 p_3\cdot p_4 + 
p_1\cdot p_4 p_2\cdot p_3 - 2 \bar M_D^2 p_2\cdot p_4 \right), 
$$
where $\bar M_D\equiv (M_D^{\rm phy}+M_{D_s}^{\rm phy})/2$, the average of the physical
masses of the $D$ and $D_s$, is introduced to match the
dimensions. The new parameters $h_{24}$ and $h_{35}$ are dimensionless, and their
relations to the old ones are $h_{24}=h_2+h_4'$ and $h_{35}=h_3+2h_5'$, where $h_4'
= h_4\bar M_D^2$ and $h_5'=h_5\bar M_D^2$.

\begin{table}
\begin{ruledtabular}
\begin{tabular}{ |l | c c c c c|}
 &$D\bar K(I=1)$ &$D\bar K(I=0)$ &$D_s K$ &$D\pi(I=3/2)$ &$D_s\pi$ 
 \\
\hline
\texttt{m007} & $-1.19(0.40)$ & $5.34(1.45)$ & $-1.58(0.14)$ & $-1.16(0.30)$ &
$0.08(0.04)$ \\
\texttt{m010} & $-1.89(0.12)$ & $6.21(1.04)$ & $-1.55(0.09)$ & $-1.38(0.10)$ &
$0.08(0.03)$ \\
\texttt{m020} & $-1.49(0.25)$ & $4.43(1.33)$ & $-1.40(0.20)$ & $-1.08(0.30)$ &
$0.13(0.05)$ \\
\texttt{m030} & $-1.59(0.13)$ & $7.46(1.56)$ & $-1.67(0.10)$ & $-1.68(0.13)$ &
$0.32(0.05)$ \\
\end{tabular}
\end{ruledtabular}
\caption{\label{Table:mf} The values of scattering lengths for five channels in lattice units.
}
\end{table}

There are four different light quark masses in our data set, corresponding to the
four ensembles (\texttt{m007}, \texttt{m010}, \texttt{m020} and \texttt{m030}) with
pion masses approximately 301~MeV, 364~MeV, 511~MeV and 617~MeV, respectively.
There are in total 20 data points in the five channels.
The values of the scattering lengths for all the channels are collected in
Table~\ref{Table:mf}.
\begin{table}
\begin{ruledtabular}
\begin{tabular}{ | l | c c c c c |}
              & $M_\pi$     & $M_K$       & $M_D$        & $M_{D_s}$    & $a$ (fm)\\
\hline
\texttt{m007} & $0.1842(7)$ & $0.3682(5)$ & $1.2081(13)$ & $1.2637(10)$ & 0.1207\\
\texttt{m010} & $0.2238(5)$ & $0.3791(5)$ & $1.2083(11)$ & $1.2635(10)$ & 0.1214 \\
\texttt{m020} & $0.3113(4)$ & $0.4058(4)$ & $1.2226(13)$ & $1.2614(12)$ & 0.1202\\
\texttt{m030} & $0.3752(5)$ & $0.4311(5)$ & $1.2320(11)$ & $1.2599(12)$ & 0.1200\\
\end{tabular}
\end{ruledtabular}
\caption{\label{Table:masses} The masses of the pion, kaon, $D$ and $D_s$ mesons
in lattice units.  
The values of the lattice spacing $a$ are also given in the last column~\cite{Bazavov:2009bb}.}
\end{table}
In order to fit to the pion mass dependence of the results, we have to express the
masses of the involved mesons in terms of the pion mass. They are the kaon, $D$ and
$D_s$ mesons, and their masses in the four ensembles are 
listed in Table~\ref{Table:masses} together with the corresponding pion masses
and values of the lattice spacing. The masses of pion and kaon are taken from Ref.~\cite{WalkerLoud:2008bp}. 
The masses of $D$ and $D_s$ mesons are from our calculations. 
The lattice spacing is set by $r_1$ in Ref.~\cite{Bazavov:2009bb}. We will use
\be
M_K= \cir{M}_K + M_\pi^2/(4 \cir{M}_K), \quad
M_D = \cir{M}_D + (h_1 + 2h_0) \frac{M_\pi^2}{\cir{M}_D},
\quad
M_{D_s} = \cir{M}_{D_s} + 2h_0 \frac{M_\pi^2}{\cir{M}_{D_s}}.
\ee
With $\cir{M}_K=551.2$~MeV, $\cir{M}_D=1942.9$~MeV, $\cir{M}_{D_s}=2062.3$~MeV
and $h_0=0.014$, the values at different pion masses shown in Table~\ref{Table:masses}
are well described. Note that, with these values, both the kaon and charmed meson 
masses at the physical pion mass are higher than their genuine physical values. For the kaon mass this is mainly due to the unphysical strange quark mass in the lattice configurations. The strange quark mass that gives the physical light pseudoscalar meson masses has been determined to be $am_s = 0.035(7)$ in Ref.~\cite{Bazavov:2009bb}, which
is lighter than the value used in our calculations. The charmed meson masses also suffer the discretization error arise both from light and charm quark actions.  The effect of the discretization errors on the masses of charmed baryons has been investigated in Ref.~\cite{Liu:2009jc}. It suggests that the discretization errors increase the singly charmed baryon masses by around 70~MeV. It is reasonable to expect that the discretization errors also increase the masses of $D$ and $D_s$. However, with the input masses of the kaon, $D$ and $D_s$ all calculated from the lattices, the fit to the scattering lengths is self-consistent.

For a pion mass
as high as 617~MeV, the kaon mass would be even higher, around 700~MeV. Such values
are too large for a controlled chiral expansion. Therefore, we will only fit to the 
ensembles \texttt{m007}, \texttt{m010} and \texttt{m020}. 
To minimize the contamination from a particular scale-setting method, we fit to the
dimensionless product of the pion mass and the scattering length. The fit was
performed using the FORTRAN package MINUIT~\cite{James:1975dr}.
\begin{table}
\begin{ruledtabular}
\begin{tabular}{|l | c c c c c c|}
Fitting range & $\chi^2/{\rm d.o.f}$ & $\tilde{a}(\lambda=1~{\rm GeV})$ & $h_{24}$  &
$h_4'$ & $h_{35}$ & $h_5'$
 \\ \hline
\texttt{m007}-\texttt{m020} & 1.06 & $-1.88^{+0.07}_{-0.09}$ &
$-0.10^{+0.05}_{-0.06}$ & $-0.32^{+0.35}_{-0.34}$ & $0.25\pm0.13$ &
$-1.88^{+0.63}_{-0.61}$ \\           
\end{tabular}
\end{ruledtabular}
\caption{\label{Table:fit6p} Results of fitting to the lattice data of the scattering
lengths with 5 parameters.
}
\end{table}
The best fit has $\chi^2/{\rm d.o.f}=1.06$, and the resulting parameters are
collected in Table~\ref{Table:fit6p}, where the asymmetric $1\sigma$ uncertainties
are calculated using the MINOS algorithm in MINUIT. 
\begin{figure}[t]
\begin{center}
\includegraphics[width = \textwidth]{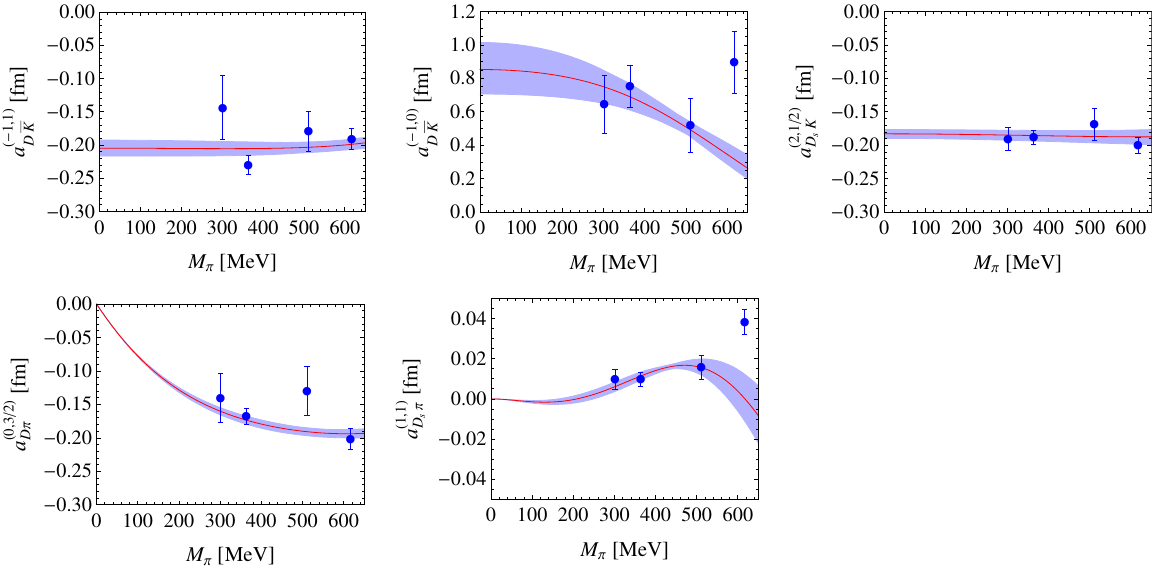}
\end{center}
\vglue-8mm
\caption{\label{Fig:fit6p3data} Fit to the data of the scattering lengths
corresponding to ensembles \texttt{m007}-\texttt{m020} in each channel. The
superscript $(S,I)$ is the (strangeness, isospin) for each channel.}
\end{figure}
A comparison of the best fit and the lattice data is shown in
Fig.~\ref{Fig:fit6p3data}, where the solid curves correspond to the results of the
best fit, and the bands reflect the uncertainties propagated from the lattice
data. At the physical pion mass, the extrapolated scattering lengths for the five
channels are presented in Table~\ref{Table:ExtrapolatedScattLen}.

One can see that all the dimensionless parameters have a natural size, 
i.e., the absolute values of $h_{24,35}$ and $h_{3,5}'$ are of
order unity. The value of $h_0=0.014$ is much smaller than $h_1=0.42$. This is
consistent with the $N_c$ counting because the $h_0$ term is suppressed by $1/N_c$ as
compared to the $h_1$ term. Furthermore, we also have the hierarchies $|h_4'|<|h_5'|$ and
$|h_2|<|h_3|$ (recall that $h_2=h_{24}-h_4'$ and $h_3=h_{35}-2h_5'$). For both cases, the
left hand sides are suppressed by $1/N_c$ as compared to the right hand sides. 

When performing the fit, we have used the physical value for the pion decay constant
$F=92.21$~MeV~\cite{PDG}.
The difference from the chiral limit value and hence its pion mass dependence is a
higher order effect, and is neglected here, although it might have some influence. 

\begin{table}
\begin{ruledtabular}
\begin{tabular}{|l | c c c c c|}
Channels & $D\bar{K} (I=1)$ & $D\bar{K}(I=0)$ & $D_sK$ & $D\pi (I=3/2)$ & $D_s\pi$ \\
\hline
 $a$~(fm)  & $-0.20(1)$  & $0.84(15)$ &
 $-0.18(1)$ & $-0.100(2)$ & $-0.002(1)$ \\
\end{tabular}
\end{ruledtabular}
\caption{\label{Table:ExtrapolatedScattLen}The scattering lengths extrapolated to 
the physical light quark masses. }
\end{table}

\section{Implications for other channels }
\label{sec:ds2317}

\subsection{Scattering lengths}

\begin{table}[tb]
\begin{ruledtabular}
\begin{tabular}{|l c c c c|}
Channels & $D\pi(I=1/2)$ & $DK(I=0)$ &$DK(I=1)$ &$D_s\bar{K}$ \\
\hline
$a$~(fm)  & $0.37^{+0.03}_{-0.02}$ & $-0.84^{+0.17}_{-0.22}$ & $0.07\pm0.03 + i
(0.17^{+0.02}_{-0.01})$ & $-0.09^{+0.06}_{-0.05} + i (0.44\pm0.05)$ \\
\end{tabular}
\end{ruledtabular}
\caption{\label{Table:PredictedScattLen} Scattering lengths of $D\pi(I=1/2)$,
$DK(I=1)$ and $D_sK$ at the physical pion mass predicted from the fit. }
\end{table}

\begin{figure}[tb]
\begin{center}
\includegraphics[width = \textwidth]{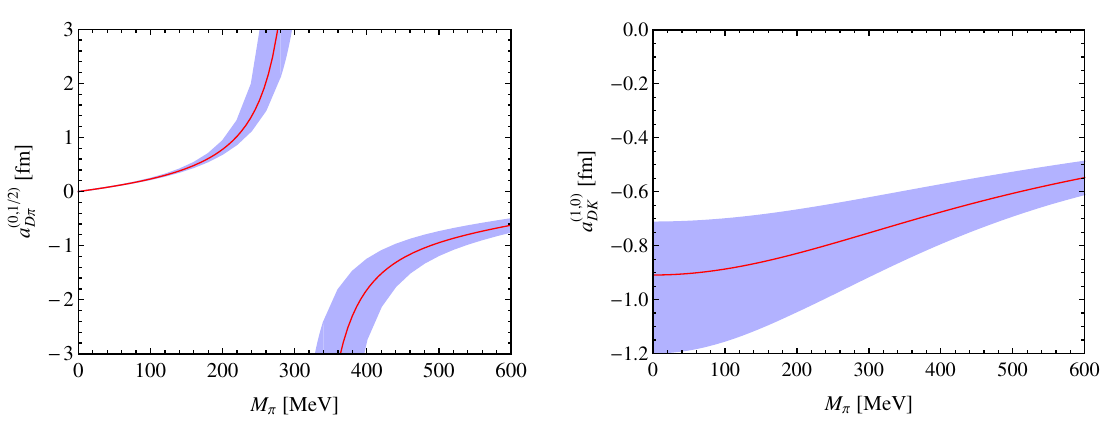}
\end{center}
\vglue-8mm
\caption{\label{Fig:dpidk6p} Predicted pion mass dependence of the $D\pi(I=1/2)$ and
$DK(I=0)$ scattering lengths using parameters from the 5-parameter fit. The solid
curves are calculated using the parameters from the best fit, and the bands reflect
the uncertainties. }
\end{figure}
In this work we did not calculate the scattering lengths on the lattice for the
channels whose Wick contractions involve disconnected diagrams due to the 
computational difficulties, as well as the additional
lattice artifacts present in these channels due to the use of Kogut-Susskind sea quarks.
However, once we have determined the LECs in the chiral Lagrangian, we can make
predictions on the scattering lengths of these channels. The results for the
scattering lengths of $D\pi(I=1/2)$, $DK(I=0)$, $DK(I=1)$ and $D_s\bar{K}$ at the physical pion mass are presented in 
Table~\ref{Table:PredictedScattLen}.
For these predictions, we have required that the masses of the involved mesons
at the physical pion mass coincide with their physical values, i.e.,  
$\cir{M}_K=485.9$~MeV, $\cir{M}_D=1862.7$~MeV, $\cir{M}_{D_s}=1968.2$~MeV are used.
For $DK(I=1)$, the imaginary part of
the scattering length originates because it couples to $D_s\pi$ with a lower
threshold. Similarly, $D_s\bar{K}$ couples to $D\pi$ and $D\eta$ so that the scattering
length is complex, too. The result for the $D\pi (I=1/2)$ channel is consistent with
the indirect extraction from lattice calculations of the $D\pi$ scalar form
factor $(0.41\pm0.06)$~fm~\cite{Flynn:2007ki}. At a pion mass of about 266~MeV, our
prediction is $2.30^{+2.40}_{-0.66}$~fm, larger than the very recent full QCD
calculation $(0.81\pm0.14)$~fm~\cite{Mohler:2012}. 
From Fig.~\ref{Fig:dpidk6p}, one sees that such a pion mass is close
to the transition point where the scattering length changes sign due to the
generation of a pole (for more discussions, see~\cite{Guo:2009ct}). In such a region, the value 
of the scattering length changes quickly. For instance, decreasing the pion mass
by 40~MeV, we would get a much smaller value $1.11^{+0.36}_{-0.17}$~fm.

The most interesting channel is the one with $(S,I)=(1,0)$, where the
$D_{s0}^*(2317)$ resides. This state was proposed to be a hadronic molecule with a
dominant $DK$ component by several
groups~\cite{Barnes:2003dj,vanBeveren:2003kd,Kolomeitsev:2003ac,Guo:2006fu,Gamermann:2006nm}.
The attraction in this channel is so strong that a pole emerges in the resummed
amplitude. Within the range of $1\sigma$ uncertainties of the the parameters, there
is always a pole on the real axis in the first Riemann sheet, which corresponds to a bound state.
If we use the physical values for all the meson masses, the pole position is
$2315^{+18}_{-28}$~MeV. The central value corresponds to the pole found using the
best fit parameters. It is very close to the observed mass of the $D_{s0}^*(2317)$, 
$(2317.8\pm0.6)$~MeV~\cite{PDG},
and it is found in the channel with the same quantum numbers as that state.
Therefore, one is encouraged to identify the bound state pole with the $D_{s0}^*(2317)$. 

As emphasized in, for instance,
Refs.~\cite{Weinberg:1965zz,Baru:2003qq}, if there is an $S$-wave shallow bound
state, the scattering length is related to the binding energy, and to the wave
function renormalization constant $Z$, with $(1-Z)$ being the probability of finding
the molecular component in the physical state (for $Z=0$, the physical state is
purely a bound state). The relation reads
\be
 a = -2 \left( \frac{1-Z}{2-Z} \right) \frac1{\sqrt{2\mu\epsilon}} \left(
 1+\mathcal{O}(\sqrt{2\mu\epsilon}/\beta) \right),
 \label{eq:weinberg}
\ee
where $\mu$ and $\epsilon$ are the reduced mass and binding energy, respectively.
Corrections of the above equation come from neglecting the range of
forces, $1/\beta$, which contains information of the $D_s\eta$ channel.
Were the
$D_{s0}^*(2317)$ a pure $DK$ bound state ($Z=0$), the value of $DK(I=0)$ scattering
length would be $a=-1.05$~fm, which coincides with the range in Table~\ref{Table:PredictedScattLen}.
From Eq.~\eqref{eq:weinberg}, the factor $Z$ is found to be in the range $[0.27,0.34]$.
This means that the main component of the pole, corresponding to the $D_{s0}^*(2317)$, 
is the $S$-wave $DK$ in the isoscalar channel.

\subsection{Isospin breaking width of the $\bm{D_{s0}^*(2317)}$}
\label{sec:newfit}

\begin{table}[tb]
\begin{ruledtabular}
\begin{tabular}{|l c c c c c|}
Fitting range & $\chi^2/{\rm d.o.f}$& $h_{24}$  &
$h_4'$ & $h_{35}$ & $h_5'$
 \\ \hline
\texttt{m007}-\texttt{m020} & 0.97 &  
$-0.10^{+0.05}_{-0.06}$ & $-0.30^{+0.31}_{-0.28}$ & $0.26^{+0.09}_{-0.10}$ &
$-1.94^{+0.46}_{-0.38}$ \\           
\end{tabular}
\end{ruledtabular}
\caption{\label{Table:fit5p} Results of fitting to the lattice data of the scattering
lengths with 4 parameters. The subtraction constant is solved from fixing the pole
in the $(S,I)=(1,0)$ channel to 2317.8~MeV. }
\end{table}
In the following, we will assume that the $D_{s0}^*(2317)$ corresponds to the pole
generated in the $(S,I)=(1,0)$ channel, and explore the implications of our lattice
calculation on this state. We will fix the pole position to the mass of the
$D_{s0}^*(2317)$, 2317.8~MeV~\cite{PDG}, on the first Riemann sheet. We fit
the lattice results of the scattering lengths with four parameters $h_{24}$, $h_{35}$, 
$h_4'$ and $h_5'$, and adjust the subtraction
constant $\tilde{a}(\lambda=1~{\rm GeV})$ to reproduce the mass of the $D_{s0}^*(2317)$.
Again, we only fit to the ensembles \texttt{m007}, \texttt{m010} and \texttt{m020}. 
The best fit gives $\chi^2/{\rm d.o.f}=0.97$, which is
slightly smaller than the one with one more parameter in Section~\ref{sec:chiralex}. The
parameter values together with the $1\sigma$ statistical
uncertainties are given in Table~\ref{Table:fit5p}. 
The parameter values are similar to the ones obtained in the 5-parameter fit, but
with smaller uncertainties. All the dimensionless LECs are of natural
size, and the $N_c$ hierarchies are the same as before.

\begin{figure}[tb]
\begin{center}
\includegraphics[width = \textwidth]{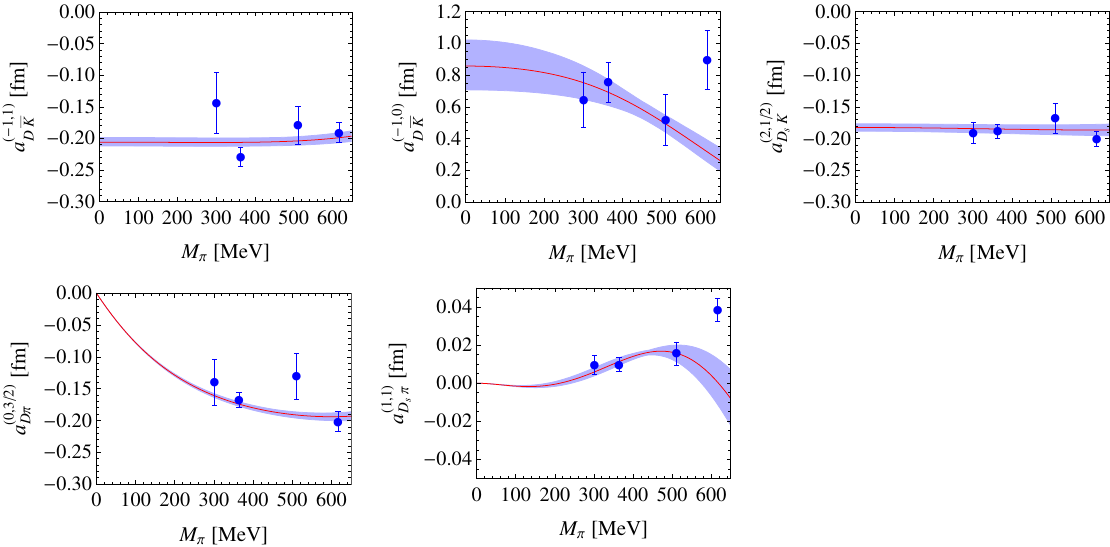}
\end{center}
\vglue-8mm
\caption{\label{Fig:fit5p} Fit to the data corresponding to ensembles
\texttt{m007}-\texttt{m020} in each channel with 4 parameters. The subtraction
constant is solved from fixing the pole in the $(S,I)=(1,0)$ channel to 2317.8~MeV. }
\end{figure}

The fitted results are presented in Fig.~\ref{Fig:fit5p}.
\begin{table}
\begin{ruledtabular}
\begin{tabular}{|l | c c c c c|}
Channels & $D\bar{K} (I=1)$ & $D\bar{K}(I=0)$ & $D_sK$ & $D\pi (I=3/2)$ & $D_s\pi$ \\
\hline
 $a({\rm fm})$  & $-0.21(1)$  & $0.84(15)$ & $-0.18(1)$ & $-0.100(1)$ &
 $-0.002(1)$ \\
\end{tabular}
\end{ruledtabular}
\caption{\label{Table:ExtrapolatedScattLen5p}The scattering lengths extrapolated to 
the physical light quark masses from the 4-parameter fit. }
\end{table}
At the physical pion mass, the extrapolated values of the scattering lengths are
listed in Table~\ref{Table:ExtrapolatedScattLen5p}.
The results are quite similar to the ones with in the last section, yet with slightly 
smaller uncertainties. 

\begin{table}
\begin{ruledtabular}
\begin{tabular}{|l | c c c c|}
Channels & $D\pi(I=1/2)$ & $DK(I=0)$ &$DK(I=1)$ &$D_s\bar{K}$ \\
\hline
$a$~(fm)  & $0.37\pm0.01$ & $-0.86\pm0.03$ & $0.04^{+0.05}_{-0.01}
+ i (0.16^{+0.02}_{-0.01})$ & $-0.06^{+0.01}_{-0.05} + i (0.45\pm0.05)$ \\
\end{tabular}
\end{ruledtabular}
\caption{\label{Table:PredictedScattLen5p} Scattering lengths of $D\pi(I=1/2)$,
$DK(I=0)$, $DK(I=1)$ and $D_sK$ at the physical pion mass predicted from the
5-parameter fit.
}
\end{table}
With the newly fitted parameters, the scattering lengths for several other channels
are predicted, and the results are listed in Table~\ref{Table:PredictedScattLen5p}. 
Again, the values are compatible with the ones in Table~\ref{Table:PredictedScattLen}.
One sees that the value for the $DK(I=0)$ channel is close to the result of Eq.~(\ref{eq:weinberg}), 
$-1.05$~fm, with $Z=0$.
The deviation from this value is partly due to the coupled-channel $D_s\eta$, and
partly due to the energy dependence in the interaction.
Using Eq.~\eqref{eq:weinberg}, the value of $Z$ is again in the range of $[0.27,0.34]$. 
Both the stability of the fit and the small $Z$ indicates that the main component of the 
$D_{s0}^*(2317)$ is the isoscalar $DK$ molecule.

\begin{figure}[t]
\begin{center}
\includegraphics[width = \textwidth]{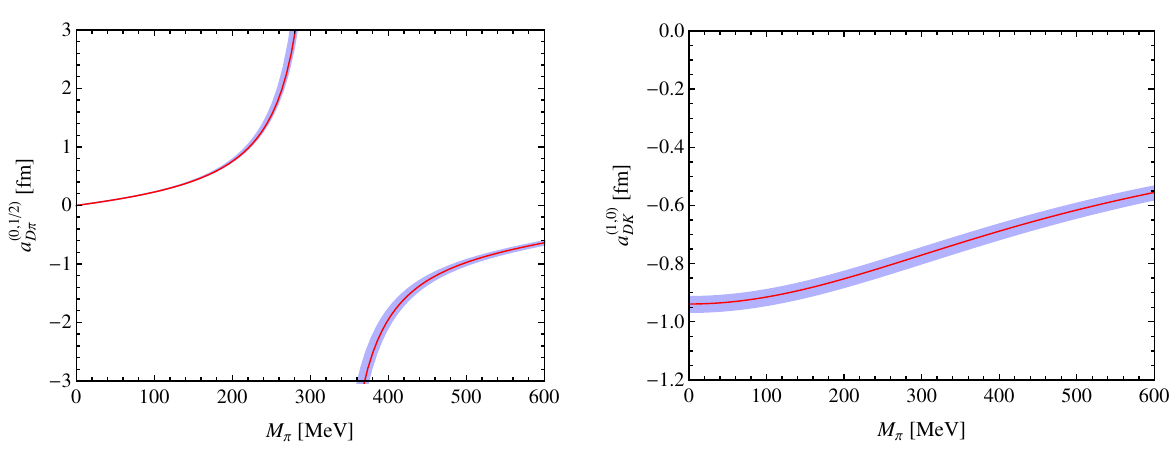}
\end{center}
\vglue-8mm
\caption{\label{Fig:dpidk5p} Predicted pion mass dependence of the $D\pi(I=1/2)$ and
$DK(I=0)$ scattering lengths using parameters from the 4-parameter fit. The solid
curves are calculated using the parameters from the best fit, and the bands reflect
the uncertainties. }
\end{figure}
We show the predictions for the pion mass dependence of the scattering lengths
for the $D\pi (I=1/2)$ and $DK(I=0)$ channels using parameters from this fit
in Fig~\ref{Fig:dpidk5p}.
The result for the $D\pi (I=1/2)$ channel at the physical pion mass 
is still consistent with the indirect extraction, $(0.41\pm0.06)$~fm, in
Ref.~\cite{Flynn:2007ki}, and the result at $M_\pi=266$~MeV,
$2.09^{+0.31}_{-0.11}$~fm, is again larger than $(0.81\pm0.14)$~fm obtained
in~\cite{Mohler:2012}. As before, in such a region,
the value of the scattering length changes quickly. For instance, decreasing the pion
mass to 220~MeV, one would get a much smaller value $(0.98^{+0.06}_{-0.03})$~fm.

All the above calculations have assumed the same mass for the
up and down quarks, and neglected the electromagnetic interaction. This is the
isospin symmetric case. However, the $D_{s0}^*(2317)$ decays into the isovector final
state $D_s\pi$. In order to calculate this isospin breaking decay width, one has to
take into account both the up and down quark mass difference and virtual photons.
This has been done in Ref.~\cite{Guo:2008gp}. In Ref.~\cite{Guo:2008gp}, the $N_c$-suppressed operators,
i.e. the $h_0$, $h_2$ and $h_4$ terms have been dropped, and a somewhat arbitrarily
chosen natural range $[-1,1]$ was taken for $h_5'$. The isospin breaking decay width
was calculated to be $\Gamma(D_{s0}^*(2317)\to D_s
\pi)=(180\pm110)$~keV~\cite{Guo:2008gp}.
With the values of all the $h_i$'s in Table~\ref{Table:fit5p}, the result is updated
to be
\be
 \Gamma(D_{s0}^*(2317)\to D_s \pi) = (133\pm19)~{\rm keV}.
\ee
We have used the isospin breaking quark mass ratio
$(m_d-m_u)/(m_s-\hat{m})=0.0299\pm0.018$, where $\hat{m}=(m_u+m_d)/2$, which
is calculated using the lattice averages (up to end of 2011) of the
light quark masses~\cite{Laiho:2009eu,latticeaverages}.

\section{Summary and discussion}

The low-energy interaction between a light pseudoscalar meson and a
charmed pseudoscalar meson was studied. We have calculated scattering lengths of five
channels $D\bar K(I=0)$, $D\bar K(I=1)$, $D_s K$, $D\pi(I=3/2)$ and $D_s\pi$ with
four ensembles. Among these channels, the interaction of $D\bar K(I=0)$ is attractive,
and that of the others is repulsive. The interaction of $D_s\pi$ is very weak, which is expected. The
$D_s\pi$  and $DK(I=1)$ channels are mixed since they have the same quantum numbers. 
To perform a more reliable analysis of these two channels, we need to construct the 
correlation matrix and use the variational method to extract the energies of the two 
channels. The chiral extrapolation was performed using SU(3) unitarized chiral
perturbation theory, and the LECs $h_i$'s in the chiral Lagrangian were determined
from a fit to the lattice results. With the same set of parameters, we made predictions
on other channels including $DK(I=0)$, $DK(I=1)$, $D\pi(I=1/2)$ and $D_s\bar K$. In
particular, we found that the attractive interaction in the $DK(I=0)$ channel is
strong enough so that a pole is generated in the unitarized scattering amplitude. 
Within $1\sigma$ uncertainties of the parameters, the pole is at $2315^{+18}_{-28}$~MeV,
and it is always below the $DK$ threshold. From calculating the wave function normalization constant,
it is found that this pole is mainly an $S$-wave $DK$ bound state.
By further fixing the pole to the observed mass of $D_{s0}^*(2317)$, we revisited the isospin breaking 
decay width of the $D_{s0}^*(2317)\to D_s\pi$. The result $(133\pm19)$~keV updates the old
result $(180\pm110)$~keV obtained in Ref.~\cite{Guo:2008gp}. It is nice to see that
the uncertainty of the width shrinks a lot. We want to stress that the width is much 
larger than the isospin breaking width of a $c\bar s$ meson, which is of order 10~keV.

It is possible to further constrain the values of $h_i$'s once simulations in
other channels are done. Although a precise calculation of the other channels
requires disconnected diagrams, one may obtain valuable information from the 
connected part only. The connected and disconnected parts can be calculated
separately using partially quenched chiral perturbation theory (for
reviews, see Refs.~\cite{Golterman:2009kw,Sharpe:2006pu}), then a fit to the lattice
calculation can be performed. This point has already been stressed, for instance, in
Ref.~\cite{DellaMorte:2010aq} for the hadronic vacuum polarization and in
Ref.~\cite{Juttner:2011ur} for the scalar form factor of the pion.

In our chiral extrapolation, the resummed chiral amplitude is of 
$\mathcal{O}(p^2)$. At this order, there is no counterterm to absorb the divergence
of the loop function $G(s)$, because loops only start from $\mathcal{O}(p^3)$. As a
result, we had to regularize the divergent loop by a subtraction constant, the
pion-mass dependence of which was neglected. Were a full one-loop calculation
available, the chiral amplitudes can be renormalized at one-loop order, and the
representation of the pion mass dependence would be improved. However, more
unknown LECs will be introduced in this way, and it is difficult to perform a fit
with all of them to the present data. As mentioned above, more data can come from
calculating the other channels, which is useful even if the disconnected contribution
is neglected. Such a study with an improved chiral extrapolation is relegated to the
future.

\section*{Acknowledgements}

We would like to thank the referee for her/his very helpful comments on the first
version of the manuscript.
We thank the NPLQCD and LHP Collaborations for sharing their light and strange
propagators. We also thank Andr\'{e} Walker-Loud and Huey-Wen Lin for important
contributions to this work.
Calculations were performed using the Chroma software suite, on computer clusters at 
Jefferson Laboratory (USQCD SciDAC supported) and the College of William and Mary 
(Cyclades cluster supported by the Jeffress Memorial Trust grant J-813). The work of
L.~L. and K.~O. is supported in part by Jefferson Science Associates under U.S. DOE
Contract No.
DE-AC05-06OR23177, and in part by  DOE grants DE-FG02-07ER41527 and
DE-FG02-04ER41302. L.~L. also acknowledges support from the European Union under Grant Agreement number 238353 (ITN STRONGnet).  The work of F.-K.~G., C.~H. and U.-G.~M. is supported in part by
the DFG and the NSFC through funds provided to the Sino-German CRC 110 ``Symmetries and
the Emergence of Structure in QCD'',  and the EU I3HP ``Study of Strongly
Interacting Matter'' under the Seventh Framework Program of
the EU. U.-G.~M. also thanks the BMBF for support (Grant
No. 06BN9006). F.-K.~G. also acknowledges partial support from
the NSFC (Grant No. 11165005).

\bibliographystyle{apsrev}
\bibliography{Scattering}

\end{document}